\documentclass[twocolumn,showpacs,amsmath,amssymb,10pt]{revtex4}
\usepackage{amsfonts}
\usepackage{bm}

\newcommand{\M}{M_2(\mathbb{C})}
\newcommand{\MM}{M_4(\mathbb{C})}

\newcommand{\ot}{{\,\otimes\,}}
\newcommand{{\Cd}}{{\mathbb{C}^d}}

\def\oper{{\mathchoice{\rm 1\mskip-4mu l}{\rm 1\mskip-4mu l}%
{\rm 1\mskip-4.5mu l}{\rm 1\mskip-5mu l}}}
\def\<{\langle}
\def\>{\rangle}

\newtheorem{proposition}{Proposition}

\newtheorem{remark}{Remark}

\newtheorem{lemma}{Lemma}

\begin{document}

\title{\textbf{Constructing optimal entanglement witnesses. II
}} \author{Dariusz
Chru\'sci\'nski and Justyna Pytel\thanks{email:
darch@fizyka.umk.pl}}
\affiliation{Institute of Physics, Nicolaus Copernicus University,\\
Grudzi\c{a}dzka 5/7, 87--100 Toru\'n, Poland}


\begin{abstract}

We provide a  class of optimal nondecomposable entanglement
witnesses for $4N \times 4N$ composite quantum systems or,
equivalently,  a new construction of nondecomposable positive maps
in the algebra of $4N \times 4N$ complex matrices. This construction
provides natural generalization of the Robertson map. It is shown
that their structural physical approximations give rise to
entanglement breaking channels.

\end{abstract}

\maketitle

\section{Introduction}

Entanglement is one of the essential features of quantum physics and
is fundamental  in modern quantum technologies \cite{QIT,HHHH}. The
most general approach to characterize quantum entanglement uses a
notion of an entanglement witness (EW) \cite{EW1,EW2}.
There is a considerable effort devoted to constructing and analyzing
the structure of EWs
\cite{Terhal2,O,Lew1,Lew2,Lew3,Bruss,Toth,Bertlmann,Brandao,O3,W-Bell,Breuer,Hall,how}.
(see also Ref. \cite{Guhne} for the recent review of entanglement
detention). However, the general method of constructing an EW is
still not known.

Due to the Choi-Jamio{\l}kowski isomorphism,   any EW in
$\mathcal{H}_A \ot \mathcal{H}_B$ corresponds to a linear positive
map $\Lambda : \mathcal{B}(\mathcal{H}_A) \rightarrow
\mathcal{B}(\mathcal{H}_B)$, where  $\mathcal{B}(\mathcal{H})$
denotes the space of bounded operators on the Hilbert space
$\mathcal{H}$. Recall that a linear map $\Lambda$ is said to be
positive if it sends a positive operator on $\mathcal{H}_A$ into a
positive operator on $\mathcal{H}_B$. It turns out \cite{EW1} that a
state $\rho$ in $\mathcal{H}_A \ot \mathcal{H}_B$ is separable iff
$(\oper_A \ot \Lambda)\rho$ is positive definite for all positive
maps $\Lambda : \mathcal{B}(\mathcal{H}_B) \rightarrow
\mathcal{B}(\mathcal{H}_A)$. Unfortunately, in spite of the
considerable effort, the structure of positive maps is rather poorly
understood (see Refs. \cite{OSID-W,CMP,PRA-Gniewko} for the recent
research).

In a recent paper  we provided a class of nondecomposable positive
maps $M_{2K}(\mathbb{C})$ ($M_d(\mathbb{C})$ denotes the algebra of
$d \times d$ complex matrices) \cite{Justyna-1}. For $K=2$ they are
closely related to the Breuer-Hall maps in $\MM$ \cite{Breuer,Hall}.
It was shown that they provide a class of optimal entanglement
witnesses.  In the present paper -- treated as a second part of Ref.
\cite{Justyna-1} -- we provide another construction  of a family of
positive maps in $M_{4N}(\mathbb{C})$ (see Ref.  \cite{Justyna-1}
for all definitions). Our construction provides a natural
generalization of the celebrated Robertson map in $\MM$
\cite{Robertson}.  We show that proposed maps are nondecomposable
(i.e., they are able to detect entangled PPT [positive partial
transposed] states) and optimal (i.e., they are able to detect the
maximal number of entangled states). As a byproduct we construct new
families of PPT entangled states detected by our maps.

The paper is organized as follows: Section~\ref{ROB} provides the
basic construction of a family of positive maps in
$M_{4N}(\mathbb{C})$. Then in Section~\ref{EW} we study the basic
properties of our maps/witnesses (nondecomposability and
optimality). In Section~\ref{SPA} we discuss the  structural
physical approximation (SPA) \cite{SPA1,SPA2,SPA3} of our maps. It
is shown that the corresponding SPA gives rise to entanglement
breaking channels and hence it supports a recent conjecture
\cite{SPA3}. Final conclusions are collected in the last Section.

\section{Generalized  Robertson maps}  \label{ROB}

Our starting point is the reduction map in the matrix algebra $\M$
\begin{equation}\label{R2-a}
    R_2(X)=\mathbb{I}_{2}\mathrm{Tr}X-X\ ,
\end{equation}
and hence its action on a matrix $X = ||x_{ij}||$ reads as follows
\begin{equation}\label{R2-b}
    \left( \begin{array}{cc} x_{11} & x_{12} \\ x_{21} & x_{22}
    \end{array} \right) \ \longrightarrow \ \left( \begin{array}{cc} x_{22} & - x_{12} \\ -x_{21} & x_{11}
    \end{array} \right)\ .
\end{equation}
It is clear that $R_2$ is a positive map, since for any rank-1
projector $P$ one finds $R_2(P) = \mathbb{I}_2 - P = P^\perp \geq
0$. There are several ways to generalize formulae (\ref{R2-a}) and
(\ref{R2-b}) for higher dimensions. An obvious generalization of
(\ref{R2-a}) reads as
\begin{equation}\label{RK}
    R_K(X)=\mathbb{I}_{K}\mathrm{Tr}X-X\ ,
\end{equation}
that is, $R_K$ is the reduction map in  $M_K(\mathbb{C})$.  The
formula (\ref{R2-b}) may be  generalized to $M_{2K}(\mathbb{C})$.
Let us observe that $M_{2K}(\mathbb{C}) = \M \ot M_{K}(\mathbb{C})$
and hence any matrix $X \in M_{2K}(\mathbb{C})$ may be represented
as
\begin{equation}\label{}
    X = \sum_{k,l=1}^2 |k\>\<l| \ot X_{kl}\ ,
\end{equation}
where $\{|1\>,|2\>\}$ denotes the standard basis in $\mathbb{C}^2$
and $X_{kl} \in M_{K}(\mathbb{C})$. In what follows we shall use the
following notation
\begin{equation}\label{BLOCK}
    X = \left( \begin{array}{c|c} X_{11} & X_{12} \\ \hline X_{21} & X_{22} \end{array}
\right)\ ,
\end{equation}
to display the block structure of $X$. Now, one has two maps in
$M_{2K}(\mathbb{C})$ that reduce to (\ref{R2-b}) for $K=1$:
\begin{equation}\label{I}
    \left( \begin{array}{c|c} X_{11} & X_{12} \\ \hline X_{21} & X_{22} \end{array}
\right) \ \longrightarrow \ \frac 1K \left( \begin{array}{c|c}
X_{22} & - X_{12} \\ \hline - X_{21} & X_{11} \end{array} \right) \
,
\end{equation}
and
\begin{equation}\label{II}
    \left( \begin{array}{c|c} X_{11} & X_{12} \\ \hline X_{21} & X_{22} \end{array}
\right) \ \longrightarrow \ \frac 1K \left( \begin{array}{c|c}
\mathbb{I}_K {\rm Tr} X_{22} & - X_{12} \\ \hline - X_{21} &
\mathbb{I}_K {\rm Tr} X_{11} \end{array} \right) \ .
\end{equation}
It is easy to show that both maps (\ref{I}) and (\ref{II}) are
decomposable and hence cannot  be used to detect PPT entangled
states.

The first example of nondecomposable positive map in
$M_{2K}(\mathbb{C})$ was provided by Robertson \cite{Robertson} for
$K=2$:
\begin{equation}\label{Phi-1}
    \Phi_4\left( \begin{array}{c|c} X_{11} & X_{12} \\ \hline X_{21} & X_{22} \end{array}
\right) = \frac 12 \left( \begin{array}{c|c} \mathbb{I}_2\,
\mbox{Tr} X_{22} &  - A_{12} \\ \hline  - A_{21} & \mathbb{I}_2\,
\mbox{Tr} X_{11}
\end{array} \right) \ ,
\end{equation}
where
\begin{eqnarray*}
  A_{12} &=&  X_{12} + R_2(X_{21}) \
\end{eqnarray*}
and
\begin{eqnarray*}
  A_{21} &=& X_{21} + R_2(X_{12}) \ .
\end{eqnarray*}
Recently,  Robertson map was generalized to $M_{2K}(\mathbb{C})$ as
\cite{Justyna-1}
\begin{equation}\label{Phi-1}
    \Psi_{2K}\left( \begin{array}{c|c} X_{11} & X_{12} \\ \hline X_{21} & X_{22} \end{array}
\right) = \frac{1}{2K} \left( \begin{array}{c|c} \mathbb{I}_K\,
\mbox{Tr} X_{22} &  - B_{12} \\ \hline  -B_{21} & \mathbb{I}_K\,
\mbox{Tr} X_{11}
\end{array} \right) \ ,
\end{equation}
where
\begin{eqnarray*}
  B_{12} &=&  X_{12} + R_N(X_{21}) \
\end{eqnarray*}
and
\begin{eqnarray*}
  B_{21} &=& X_{21} + R_N(X_{12}) \ ,
\end{eqnarray*}
and it was proved that $\Psi_{2K}$ is nondecomposable. In the
present paper we propose another generalization of $\Phi_4$ for
$M_{4N}(\mathbb{C})$.  Let us observe that
\begin{equation}\label{}
    R_2(X) = \sigma_y X^{\rm T} \sigma_y \ ,
\end{equation}
where $\sigma_y$ stands for the $y$--Pauli matrix. What is important
is that $\sigma_y$ is unitary and anti-symmetric. Essentially (up to
a phase factor), it is the only  antisymmetric unitary matrix in
$\M$. Now, let us define the following map in $M_{4N}(\mathbb{C})$:
\begin{equation}\label{Phi-1}
    \Phi^U_{4N}\left( \begin{array}{c|c} X_{11} & X_{12} \\ \hline X_{21} & X_{22} \end{array}
\right) = \frac{1}{2N} \left( \begin{array}{c|c} \mathbb{I}_{2N}\,
\mbox{Tr} X_{22} &  - C^U_{12} \\ \hline  -C^U_{21} &
\mathbb{I}_{2N}\, \mbox{Tr} X_{11}
\end{array} \right) \ ,
\end{equation}
where
\begin{eqnarray*}
  C^U_{12} &=&  X_{12} + UX_{21}^{\rm T}U^\dagger
\end{eqnarray*}
and
\begin{eqnarray*}
  C^U_{21} &=& X_{21} + UX_{12}^{\rm T} U^\dagger \ ,
\end{eqnarray*}
and $U$ is an arbitrary  antisymmetric unitary matrix in
$M_{2N}(\mathbb{C})$. The above formulae guarantee that $\Psi_{2K}$
and $\Phi^U_{4N}$ are unital, i.e.
\begin{equation}\label{}
    \Psi_{2K}(\mathbb{I}_{2K}) = \mathbb{I}_{2K}\ , \ \ \ \   \Phi^U_{4N}(\mathbb{I}_{4N}) = \mathbb{I}_{4N}\
    .
\end{equation}
Clearly, $\Psi_{2K}$ and $\Phi^U_{4N}$ coincide iff $2K=4N=4$. In
this case $U = e^{i\lambda} \sigma_y$. However, if $2K=4N > 4$, they
are different. It follows from the fact that for $K>1$, the
reduction map $R_{2K}(X)$ can not be represented as $U X^{\rm T}
U^\dagger$, with a unitary, antisymmetric $U$. Indeed, one has
$R_{2K}(|1\>\<1|) = \mathbb{I}_{2K} - |1\>\<1|$, and hence ${\rm
Tr}[R_{2K}(|1\>\<1|)]=2K-1$. On the other hand ${\rm Tr}[U |1\>\<1|
U^\dagger] = 1$. Hence, necessarily  $K=1$.

\begin{proposition}  $\Phi^U_{4N}$ defines a linear  positive map in
$M_{4N}(\mathbb{C})$.
\end{proposition}
{\bf Proof}: to prove that $\Phi^U_{4N}$ defines a positive map it
is enough to show that each rank-1 projector $P \in \MM$ is mapped
via $\Phi^U_{4N}$ into a positive element in $\MM$, that is,
$\Phi^U_{4N}(P)\geq 0$. Let $P = |\psi\>\<\psi|$ with arbitrary
$\psi$ from $\mathbb{C}^{4N}$ satisfying $\<\psi|\psi\>=1$. Now,
since $\mathbb{C}^{4N} = \mathbb{C}^{2N} \oplus \mathbb{C}^{2N}$ one
has
\begin{equation}\label{}
    \psi = \sqrt{a}\,\psi_1 \oplus \sqrt{1-a}\, \psi_2\ ,
\end{equation}
with normalized $\psi_1,\psi_2 \in\mathbb{C}^{2N}$ and $a \in
[0,1]$. One has
\begin{equation*}\label{}
    P = \left( \begin{array}{c|c} X_{11} & X_{12} \\ \hline X_{21} & X_{22} \end{array}
\right) = \left( \begin{array}{c|c} a\,|\psi_1\>\<\psi_1| & b\,|\psi_1\>\<\psi_2| \\
\hline b\,|\psi_2\>\<\psi_1| & (1-a)\,|\psi_2\>\<\psi_2|
\end{array}      \right)   \  ,
\end{equation*}
where $b =  \sqrt{a(1-a)}\,$. Therefore
\begin{equation}\label{Phi-1}
    \Phi^U_{4N}(P) = \frac{1}{2N} \left( \begin{array}{c|c} (1-a)\,\mathbb{I}_{2N} &
      - b\, M \\ \hline  - b\, M^\dagger & a\,\mathbb{I}_{2N}
\end{array} \right) \ ,
\end{equation}
where
\begin{equation}\label{}
    M = |\psi_1\>\<\psi_2| + U (|\psi_2\>\<\psi_1|)^{\rm T}
    U^\dagger\ .
\end{equation}
It is clear that if $a=0$, then
\begin{equation}\label{Phi-1}
    \Phi^U_{4N}(P) = \frac{1}{2N} \left( \begin{array}{c|c} \mathbb{I}_{2N}
    & \mathbb{O}_{2N} \\ \hline  \mathbb{O}_{2N} & \mathbb{O}_{2N} \end{array} \right)
    \geq 0 \ .
\end{equation}
Similarly, for $a=1$ one finds
\begin{equation}\label{Phi-1}
    \Phi^U_{4N}(P) = \frac{1}{2N} \left( \begin{array}{c|c} \mathbb{O}_{2N}
    & \mathbb{O}_{2N} \\ \hline  \mathbb{O}_{2N} & \mathbb{I}_{2N} \end{array} \right)
    \geq 0 \ .
\end{equation}
Assume now that $0 < a < 1$. Let us recall \cite{Bhatia} that a
Hermitian matrix $X \in M_{2K}(\mathbb{C})$,
\begin{equation*}
X = \left( \begin{array}{c|c} A & M \\ \hline M^\dagger & B
\end{array} \right)\
\end{equation*}
with strictly positive matrices $A, B \in M_K(\mathbb{C})$, is
positive if and only if
\begin{equation}\label{}
    A \geq M B^{-1} M^\dagger \ .
\end{equation}
Hence, to show that $\Phi^U_{4N}(P) \geq 0$ one has to prove
\begin{equation}\label{}
 \mathbb{I}_{2N} \geq  MM^\dagger\ .
\end{equation}
Taking into account that $(|\psi_2\>\<\psi_1|)^{\rm T} =
|\psi_1^*\>\<\psi_2^*|\,$, and $\<\psi|U|\psi^*\> = 0$ for any
unitary anti-symmetric matrix $U$, one obtains
\begin{equation}\label{}
    MM^\dagger =  Q + Q^U \ , 
\end{equation}
where $Q = |\psi_1\>\<\psi_1|$ and $Q^U = UQ^{\rm T}U^\dagger$.
Clearly, $Q$ and $Q^U$  are mutually orthogonal rank-1 projectors
and hence $Q + Q^U \leq \mathbb{I}_{2N}\,$, which proves the
positivity of $\Phi^U_{4N}$.

\begin{remark}
{\em One may replace the  antisymmetric unitary matrix $U$ by any
antisymmetric matrix satisfying $UU^\dagger \leq \mathbb{I}_{4N}$.
In particular, if $U=\mathbb{O}_{4N}$, one reproduces (\ref{II}).  }
\end{remark}

\begin{remark}
{\em Note that
\begin{equation}\label{U0}
    U_0 = \sigma_y \oplus \ldots \oplus \sigma_y \in M_{2N}(\mathbb{C})\
\end{equation}
is evidently antisymmetric and unitary. One may call $\Phi^{0}_{4N}$
corresponding to $U=U_0$ the canonical generalization of the
Robertson map. Note that if $V \in M_{2N}(\mathbb{R})$ is
orthogonal, i.e. $VV^{\rm T} = \mathbb{I}_{2N}$, then $U =
VU_0V^{\rm T}$ is antisymmetric and unitary. }
\end{remark}

\begin{remark}
{\em Let us recall that Breuer-Hall maps
\begin{equation}\label{}
    \Lambda^U_{2K}(X) = R_{2K}(X) - U X^{\rm T} U^\dagger\ ,
\end{equation}
with $U$ antisymmetric unitary matrix in $M_{2K}(\mathbb{C})$,
provide another generalization of the Robertson map. One has $\Phi_4
= \Lambda^0_4$, where again $\Lambda^0_4$ corresponds to $U=U_0$. We
stress, however, that for $K>2$, Breuer-Hall maps $\Lambda^U_{2K}$
differ both form $\Psi_{2K}$ and $\Phi^U_{4N}$.

 }
\end{remark}

\section{Entanglement witnesses}  \label{EW}

To show that a positive map $\Phi^U_{4N}$ can be used to detect
quantum entanglement one has to show that it is not completely
positive. It means that the corresponding Choi matrix
\begin{equation}\label{}
    W^U_{4N} = (\oper \ot \Phi^U_{4N}) P^+_{4N}\ ,
\end{equation}
where $P^+_d$ stands for the maximally entangled state in
$\mathbb{C}^d \ot \mathbb{C}^d$,  is not positive, i.e., it possess
a strictly negative eigenvalue. Direct calculation shows that the
spectrum of $W$ reads as follows:
\[  \frac{1}{4N} \times  \left\{   \begin{array}{cl}
-1 & \ \ \mbox{single} \\
0 & \  (12N^2 - 2)\mbox{--fold} \\
\frac 1N & \ \ 4N^2\mbox{--fold} \\
1 & \ \ \mbox{single} \end{array} \right. \ . \]
It proves that $W$ is indeed an entanglement witness.

\begin{proposition}
$W$ is a nondecomposable entanglement witness.
\end{proposition}
{\bf Proof}: to prove nondecomposability of $W$ one has to show that
there exists a PPT state $\rho$ such that ${\rm Tr}(W \rho) < 0$.
Let us construct the following density matrix
\begin{equation}\label{}
    \rho = \mathcal{N}\,\sum_{i,j=1}^{4N} |i\>\<j| \otimes \rho_{ij}\ ,
\end{equation}
where the blocks $\rho_{ij} \in M_{4N}(\mathbb{C})$ are defined as
follows: the diagonal blocks
\begin{equation}\label{}
    \rho_{ii}=\left(\begin{array}{c|c}
4N\cdot\mathbb{I}_{2N} & \mathbb{O}_{2N}\\
\hline \mathbb{O}_{2N} & \mathbb{I}_{2N}\end{array}\right)\ ,
\end{equation}
for $i=1,\ldots,2N\,$, and
\begin{equation}\label{}
    \rho_{ii}=\left(\begin{array}{c|c}
\mathbb{I}_{2N} & \mathbb{O}_{2N}\\
\hline \mathbb{O}_{2N} & 4N\cdot \mathbb{I}_{2N}\end{array}\right)\
,
\end{equation}
for $i=2N+1,\ldots,4N$. The off-diagonal blocks
\begin{equation}\label{}
\rho_{i,i+2N} = -8N^2\cdot\text{W}_{i,i+2N}\ ,
\end{equation}
for $i=1,\ldots,2N\,$. Finally, for any $i=1,\ldots,2N$ and $j
=2N+1,\ldots,4N$, provided that $j \neq i+2N$ one defines
\begin{equation}\label{}
    \rho_{ij} = |i\>\<j|\ .
\end{equation}
All the remaining elements do vanish, i.e. $\rho_{ij} =
\mathbb{O}_{2N}$. One finds for the normalization factor
\begin{equation}\label{}
    \mathcal{N} = \frac{1}{8N^2(1+4N)}\ .
\end{equation}
Direct calculation shows that $\rho \geq 0$ and $\rho^\Gamma \geq
0$, i.e., $\rho$ is PPT. Finally, one easily finds for the trace
\begin{equation}\label{}
    {\rm Tr}(W \rho) = - \frac{\mathcal{N} }{8N^2} \ ,
\end{equation}
which proves nondecomposability of $W$.

\begin{proposition}
$W$ is an optimal entanglement witness.
\end{proposition}
{\bf Proof}: to show that $W_{4N}^U$ is optimal we use the following
result of Lewenstein et al. \cite{Lew1}: if the family of product
vectors $\psi \ot \phi \in \mathbb{C}^{4N} \ot \mathbb{C}^{4N}$
satisfying
\begin{equation}\label{}
    \< \psi \ot \phi| W|\psi \ot \phi \> = 0 \ ,
\end{equation}
span the total Hilbert space $\mathbb{C}^{4N} \ot \mathbb{C}^{4N}$,
then $W$ is optimal. Let us introduce the following sets of vectors:
\begin{eqnarray*}
f_{mn} = e_m + e_n \ , \ \ \ \
g_{mn} = e_m + i e_n \ ,
\end{eqnarray*}
for each $1\leq m < n \leq 4N$. It is easy to check that $(4N)^2$
vectors $\psi_\alpha \ot \psi^*_\alpha$ with $\psi_\alpha$ belonging
to the set $ \{\, e_l\, , f_{mn}\, , g_{mn} \, \}\,$,  are linearly
independent and hence they do span $\mathbb{C}^{4N} \ot
\mathbb{C}^{4N}$. Direct calculation shows that
\begin{equation}\label{}
    \< \psi_\alpha \ot \psi^*_\alpha| W^U_{4N}|\psi_\alpha \ot \psi^*_\alpha \> = 0 \ ,
\end{equation}
which proves that $W_{4N}^U$ is an optimal EW.

\begin{remark}{\em
Actually, $W^U_{4N}$ is not only an optimal EW but even nd-optimal.
An EW $W$ is optimal if $W- A$ is not EW for any $A \geq 0$, that
is, subtracting from $W$ any positive operator one destroys
block-positivity of $W$. Now, $W$ is nd-optimal if $W- D$ is not EW
for any decomposable operator $D$ ($D$ is decomposable if $D = A +
B^\Gamma$, with $A,B\geq 0$). Clearly, any nd-optimal EW is optimal
and hence nd-optimal EWs define a proper subset of optimal
witnesses. Recall, that a nondecomposable EW $W$ is nd--optimal if
and only if both $W$ and $W^\Gamma$ are optimal. Note that
$(W^U_{4N})^\Gamma = V W^U_{4N} V^\dagger$, where the unitary matrix
$V$ is defined as follows
\begin{equation}\label{}
    V = |1\>\<1| \ot U^\dagger + |2\>\<2| \ot U \ ,
\end{equation}
and hence the optimality of $(W^U_{4N})^\Gamma $ easily follows from
the optimality of $W^U_{4N}$. }
\end{remark}

\begin{remark}
{\em Let us observe that for any unitarities $V_1,V_2 :
\mathbb{C}^{4N} \rightarrow \mathbb{C}^{4N}$ a new map
\begin{equation}\label{}
    \Phi^{U,V_1,V_2}_{4N}(X) := V_1^\dagger\Big[ \Phi^U_{4N}(V_2 X
    V_2^\dagger)\Big]V_1\ ,
\end{equation}
is again positive (unital) and nondecomposable. Indeed, positivity
is clear, and indecomposability follows from the following
observation: if $\Phi^U_{4N}$ detects a PPT entangled state
$\rho\,$, i.e., $(\oper \ot \Phi^U_{4N}) \rho \ngeq 0$, then
$\Phi^{U,V_1,V_2}_{4N}$ detects a PPT state $\widetilde{\rho} =
(\mathbb{I}_{4N} \ot V_2^\dagger) \rho (\mathbb{I}_{4N} \ot V_2)$.

The corresponding entanglement witness $W^{U,V_1,V_2}_{4N}$ reads as
follows
\begin{eqnarray}\label{}
W^{U,V_1,V_2}_{4N} &=& (\oper \ot \Phi^{U,V_1,V_2}_{4N}) P^+_{4N}
 \\ &=& \frac{1}{4N} \sum_{k,l=1}^{4N} |k\>\<l| \ot
V_1^\dagger\Big[ \Phi^U_{4N}(V_2 |k\>\<l|  V_2^\dagger)\Big]V_1\ ,
\nonumber
\end{eqnarray}
that is,
\begin{eqnarray*}\label{}
W^{U,V_1,V_2}_{4N} &=& (\mathbb{I}_{4N} \ot V_1^\dagger) \Big[(\oper
\ot \Phi^{U}_{4N}) \widetilde{P}^+_{4N} \Big] (\mathbb{I}_{4n} \ot
V_1) \ ,
\end{eqnarray*}
where
\begin{equation}\label{}
\widetilde{P}^+_{4N} = (\mathbb{I}_{4N} \ot V_2) P^+_{4N}
(\mathbb{I}_{4n} \ot V_2) \ .
\end{equation}
Using the fact that $P^+_{4N}$ is $V \ot \overline{V}$--invariant,
one obtains
\begin{equation}\label{}
W^{U,V_1,V_2}_{4N} = (\overline{V}_2^\dagger \ot V_1^\dagger)
W^U_{4N} (\overline{V}_2 \ot V_1)\ .
\end{equation}
Hence, if $\< \phi_k \ot \psi_k|W^U_{4N}|\phi_k \ot \psi_k\>=0$ and
$\phi_k \ot \psi_k$ do span $\mathbb{C}^{4N} \ot \mathbb{C}^{4N}$,
then $\< \widetilde{\phi}_k \ot
\widetilde{\psi}_k|W^U_{4N}|\widetilde{\phi}_k \ot
\widetilde{\psi}_k\>=0$, with
\begin{equation*}
\widetilde{\phi}_k \ot \widetilde{\psi}_k = (\overline{V}_2^\dagger
\ot V_1^\dagger)(\phi_k \ot {\psi}_k)\ .
\end{equation*}
Clearly, $\widetilde{\phi}_k \ot \widetilde{\psi}_k$ do span
$\mathbb{C}^{4N} \ot \mathbb{C}^{4N}$. Hence, it proves that
$W^{U,V_1,V_2}_{4N}$ defines an optimal entanglement witness.

}
\end{remark}

\section{Structural physical approximation}  \label{SPA}

 The idea of  the structural physical approximation
(SPA) \cite{SPA1,SPA2} consists of mixing a positive map $\Lambda$
with some completely positive map making the mixture
$\widetilde{\Lambda}$ completely positive. In the recent paper Ref.
\cite{SPA3},  the authors analyze the SPA to a positive map $\Lambda
: \mathcal{B}(\mathcal{\mathcal{H}_A}) \rightarrow
\mathcal{B}(\mathcal{\mathcal{H}_B})$ obtained through minimal
admixing of white noise
\begin{equation}\label{}
    \widetilde{\Lambda}(\rho) = p \frac{\mathbb{I}_B}{d_B} \, {\rm
    Tr}(\rho) + (1-p) \Lambda(\rho)\ .
\end{equation}
The minimal means that the positive mixing parameter $0 < p < 1$ is
the smallest one for which the resulting map $\widetilde{\Lambda}$
is completely positive, i.e., it defines a quantum channel.
Equivalently, one may introduce the SPA of an entanglement witness
$W$:
\begin{equation}\label{}
    \widetilde{W} = \frac{p}{d_A d_B} \mathbb{I}_A \ot \mathbb{I}_B
    + (1-p) W \ ,
\end{equation}
where $p$ is the smallest parameter for which $\widetilde{W}$ is a
positive operator in $\mathcal{H}_A \ot \mathcal{H}_B$, i.e. it
defines a (possibly unnormalized) state.

It was conjectured  that the SPA to optimal positive maps correspond
to entanglement breaking maps (quantum channels) \cite{SPA3}.
Equivalently, the SPA to optimal entanglement witnesses corresponds
to separable (unnormalized) states. We show that the family of
optimal maps/witnesses constructed in this paper  supports this
conjecture.

The corresponding SPA of $W^U_{4N}$ is therefore given by
\begin{equation}\label{}
    \widetilde{W}^U_{4N} = \frac{p}{(4N)^2}\, \mathbb{I}_{4N} \ot
    \mathbb{I}_{4N}     + (1-p)\, W^U_{4N} \ .
\end{equation}
The above definition guarantees  that ${\rm Tr} \widetilde{W}^U_{4N}
= 1$.  Using the fact that the negative eigenvalue of $W_{4N}^U$
equals ``$-1/4N$'' one easily finds the following condition for the
positivity of $\widetilde{W}^U_{4N}$
\begin{equation}\label{p}
    p \geq \frac{4N}{4N+1}\ .
\end{equation}
To show that the SPA of $\Phi^U_{4N}$ is entanglement breaking we
use the following result \cite{Justyna-1}: let $\Lambda :
M_d(\mathbb{C}) \rightarrow M_d(\mathbb{C})$ be a positive unital
map. Then the SPA of $\Lambda$ is entanglement breaking if $\Lambda$
detects all entangled isotropic states in $\mathbb{C}^d \ot
\mathbb{C}^d$. If, in addition, $\Lambda$ is self-dual, i.e.,
\begin{equation}\label{self}
    {\rm Tr}(X \cdot \Lambda(Y)) = {\rm Tr}(\Lambda(X) \cdot Y)\ ,
\end{equation}
for all $A,B \in M_d(\mathbb{C})$, then it is enough to check
whether all entangled isotropic states are detected by the
corresponding witness $W_\Lambda = (\oper \ot \Lambda)P^+_d$.

\begin{lemma} $\Phi^U_{4N}$ is self-dual.
\end{lemma}
Using the definition of $\Phi^U_{4N}$ one obtains
\begin{eqnarray*}
  {\rm Tr}[X \cdot \Phi^U_{4N}(Y)] = a - b \ ,   \\
\end{eqnarray*}
where
\begin{equation*}\label{}
    a = {\rm Tr}[X_{11}Y_{11} - X_{12}Y_{21} - X_{21}Y_{12} +
    X_{22}Y_{22} ] \ ,
\end{equation*}
and
\begin{equation*}\label{}
    b = {\rm Tr}[X_{12}UY^{\rm T}_{12}U^\dagger  + X_{21}UY_{21}^{\rm T}U^\dagger] \
    .
\end{equation*}
On the other hand,
\begin{eqnarray*}
  {\rm Tr}[\Phi^U_{4N}(X) \cdot Y] = a - b' \ ,   \\
\end{eqnarray*}
where
\begin{equation*}\label{}
    b' = {\rm Tr}[UX^{\rm T}_{12}U^\dagger Y_{12}  + UX_{21}^{\rm T}U^\dagger Y_{21}] \
    .
\end{equation*}
Now, using ${\rm Tr} X^{\rm T} = {\rm Tr} X$, and $U^{\rm T} = -U$,
one proves that $b=b'$ and hence  $\Phi^U_{4N}$ is self-dual.

Let
\begin{equation}\label{}
    \rho_\lambda = \frac{\lambda}{(4N)^2}\, \mathbb{I}_d \ot \mathbb{I}_d + (1-\lambda)
    P^+_{4N}\ ,
\end{equation}
be an isotropic state which is known to be entangled iff
\begin{equation}\label{p-iso}
\lambda < \frac{4N}{4N+1}\ .
\end{equation}

\begin{lemma}
If  $\rho_\lambda$ is entangled, then ${\rm Tr}(W^U_{4N}\cdot
\rho_\lambda) < 0$.
\end{lemma}
One has
\begin{equation}\label{}
{\rm Tr}(W^U_{4N}\cdot \rho_\lambda) = \frac{\lambda}{(4N)^2} +
(1-\lambda)\, {\rm Tr}(W^U_{4N} \cdot P^+_{4N})\ ,
\end{equation}
where we have used ${\rm Tr}W^U_{4N}=1$. Moreover,
\begin{equation*}
{\rm Tr}(W^U_{4N} \cdot P^+_{4N}) = \frac{1}{(4N)^2}
\sum_{k,l=1}^{4N} \<k| \,\Phi^U_{4N}( |l\>\<k|)\,|l\> \ .
\end{equation*}
Finally, direct calculation shows that
\begin{equation}\label{}
\sum_{k,l=1}^{4N} \<k|\, \Phi^U_{4N}( |l\>\<k|)\,|l\>  = -4N\ ,
\end{equation}
and hence
\begin{equation}\label{}
{\rm Tr}(W^U_{4N}\cdot \rho_\lambda) = \frac{1}{4N} \left(
\frac{\lambda}{4N} + \lambda -1\right)\ .
\end{equation}
Therefore, if $\lambda < 4N/(4N+1)$, then ${\rm Tr}(W^U_{4N}
\rho_\lambda)<0\,$, which shows that $W^U_{4N}$ detects all
entangled isotropic states.

\begin{remark}
{\em One easily shows that the SPA for $\Phi^{U,V_1,V_2}_{4N}$
provides again an entanglement breaking channel. }
\end{remark}

\section{Conclusions}

We have provided a new construction of EWs in $\mathbb{C}^{4N} \ot
\mathbb{C}^{4N}$. It was shown that these EWs are nondecomposable,
i.e., they are able to detect PPT entangled states. The crucial
property of witnesses $W^U_{4N}$ is optimality. Equivalently, our
construction gives rise to the new class of positive maps in
algebras of $4N \times 4N$ complex matrices. For $N=1$ this
construction reproduces the  Robertson map \cite{Robertson} and
hence it defines the special case of Brauer-Hall maps
\cite{Breuer,Hall}.

Interestingly, a class of EWs $W^U_{4N}$ is nd-optimal, i.e., both
$W^U_{4N}$ and its partial transposition $(W^U_{4N})^\Gamma$ are
optimal EWs and hence provide the best ``detectors" of  PPT
entangled states. We have shown that the structural physical
approximation for our new class of positive maps gives rise to
entanglement breaking channels and hence it supports the conjecture
of Ref. \cite{SPA3}.

\section*{Acknowledgement}
J.P.  thanks Spirydon  Michalakis for valuable discussions and kind
hospitality at Los Alamos National Laboratory. This work was
partially supported by the Polish Ministry of Science and Higher
Education Grant No 3004/B/H03/2007/33.

\end{document}